\documentclass[aps, twocolumn, prd, longbibliography, showkeys]{revtex4-2}

\usepackage{amsmath}
\usepackage{amssymb}
\usepackage{braket}
\usepackage{bm}
\usepackage{cuted}
\usepackage{dcolumn}
\usepackage{diagbox}
\usepackage{epsfig}
\usepackage{esint}
\usepackage{flushend}
\usepackage{graphicx}
\usepackage{mathrsfs}
\usepackage{mathtools}
\usepackage{newtxmath}
\usepackage{newtxtext}
\usepackage{slashed}
\usepackage{subfigure}
\usepackage[mathlines]{lineno}     
\usepackage[usenames, dvipsnames]{color}
\usepackage[pagebackref=false, colorlinks=true]{hyperref}
\definecolor{reddish}{rgb}{0.7,0.2,0.0}  
\definecolor{blueish}{rgb}{0.1,0.1,1}
\hypersetup{
linkcolor=reddish,     
citecolor=blueish,     
filecolor=blue,        
urlcolor=magenta}      

\newcommand{\rg}{r_{\rm g}}
\newcommand{\etaout}{\eta}
\newcommand{\OmegaH}{\Omega_{\rm H}}
\newcommand{\phibh}{\phi_{\rm H}}

\def\araa{ARA\&A}                
\def\apj{ApJ}                    
\def\apjl{ApJL}                  
\def\apjs{ApJS}                  

\def\aap{A\&A}                   
\def\mnras{MNRAS}                
 
\def\prd{Phys.~Rev.~D}           
\def\prl{Phys.~Rev.~Lett.}       

\begin{document}


\title{Energy Extraction from Spinning Stringy Black Holes}
\author{Koushik Chatterjee,$^{1,2}$}
\email{koushik.chatterjee@cfa.harvard.edu}
\author{Prashant Kocherlakota,$^{1,2}$}
\email{prashant.kocherlakota@cfa.harvard.edu}
\author{Ziri Younsi,$^{3}$}
\email{z.younsi@ucl.ac.uk}
\author{Ramesh Narayan$^{2,1}$}
\email{rnarayan@cfa.harvard.edu}

\affiliation{
$^1$Black Hole Initiative at Harvard University, 20 Garden St., Cambridge, MA 02138, USA\\
$^2$Center for Astrophysics, Harvard \& Smithsonian, 60 Garden St., Cambridge, MA 02138, USA\\
$^3$Mullard Space Science Laboratory, University College London, Holmbury St.~Mary, Dorking, Surrey, RH5 6NT
}

\begin{abstract}
We perform the first numerical simulations modeling the inflow and outflow of magnetized plasma in the Kerr-Sen spacetime, which describes classical spinning black holes (BHs) in string theory. We find that the Blandford-Znajek (BZ) mechanism, which is believed to power astrophysical relativistic outflows or ``jets'', is valid even for BHs in an alternate theory of gravity, including near the extremal limit. The BZ mechanism releases outward Poynting-flux-dominated plasma as frame-dragging forces magnetic field lines to twist. However, for nonspinning BHs, where the frame-dragging is absent, we find an alternate powering mechanism through the release of gravitational potential energy during accretion. Outflows from non-spinning stringy BHs can be approximately $250\%$ more powerful as compared to Schwarzschild BHs, due to their relatively smaller event horizon sizes and, thus, higher curvatures. Finally, by constructing the first synthetic images of near-extremal non-Kerr BHs from time-dependent simulations, we find that these can be ruled out by horizon-scale interferometric images of accreting supermassive BHs. 
\end{abstract}

\maketitle


\section{\label{sec:intro}Introduction}

With the first detections of gravitational waves produced from binary black hole (BH) mergers \cite{Abbott+2016a} and the first images of the supermassive compact objects M87$^*$ and Sgr A$^*$ \cite{EHTC+2019a, EHTC+2022a}, experimental gravitation has firmly entered the near-horizon regime \cite{Abbott+2016b, Abbott+2018a, Abbott+2018b, Abbott+2019a, Abbott+2019b, EHTC+2019f, Psaltis+2020, Abbott+2021a, Abbott+2021b, Abbott+2021c, Kocherlakota+2021, Volkel+2021, Abbott+2021d, EHTC+2022f, Vagnozzi+2022}.
The search for possible gravitational signatures originating on cosmological scales imprinted in such measurements is well underway \cite{Abbott+2018b, Abbott+2023}.
Novel tests that probe, e.g., the existence of new fundamental fields to describe dark matter or due to physics in the quantum, high-energy realms are also becoming possible \cite{Cardoso+2017, Hirschmann+2017, Giddings+2018, Mizuno+2018, Abbott+2018b, Abbott+2021c, Fromm+2021, Kocherlakota+2021, Volkel+2021, Abbott+2022, EHTC+2022f, Vagnozzi+2022, Volkel+2022, Franchini+2023, Roder+2023}. 

In response to this impetus from observations, numerical simulations are becoming increasingly more important as a tool for interpreting these observations.
For example, the Event Horizon Telescope (EHT) Collaboration constructed a library of 3D general-relativistic magneto-hydrodynamics (GRMHD) simulations of hot, magnetized accretion flows in Kerr BH spacetimes to enable the interpretation of the observed horizon-scale images \cite{EHTC+2019e, EHTC+2022e}.
The Kerr metric describes a stationary, axisymmetric, vacuum, spinning BH in general relativity (GR), and the comparison of synthetic images produced from these GRMHD simulations with actual observed images established a successful null test of GR \cite{EHTC+2019f, EHTC+2022f}. 
\begin{figure*}
    \centering

    \includegraphics[width=2.02\columnwidth]{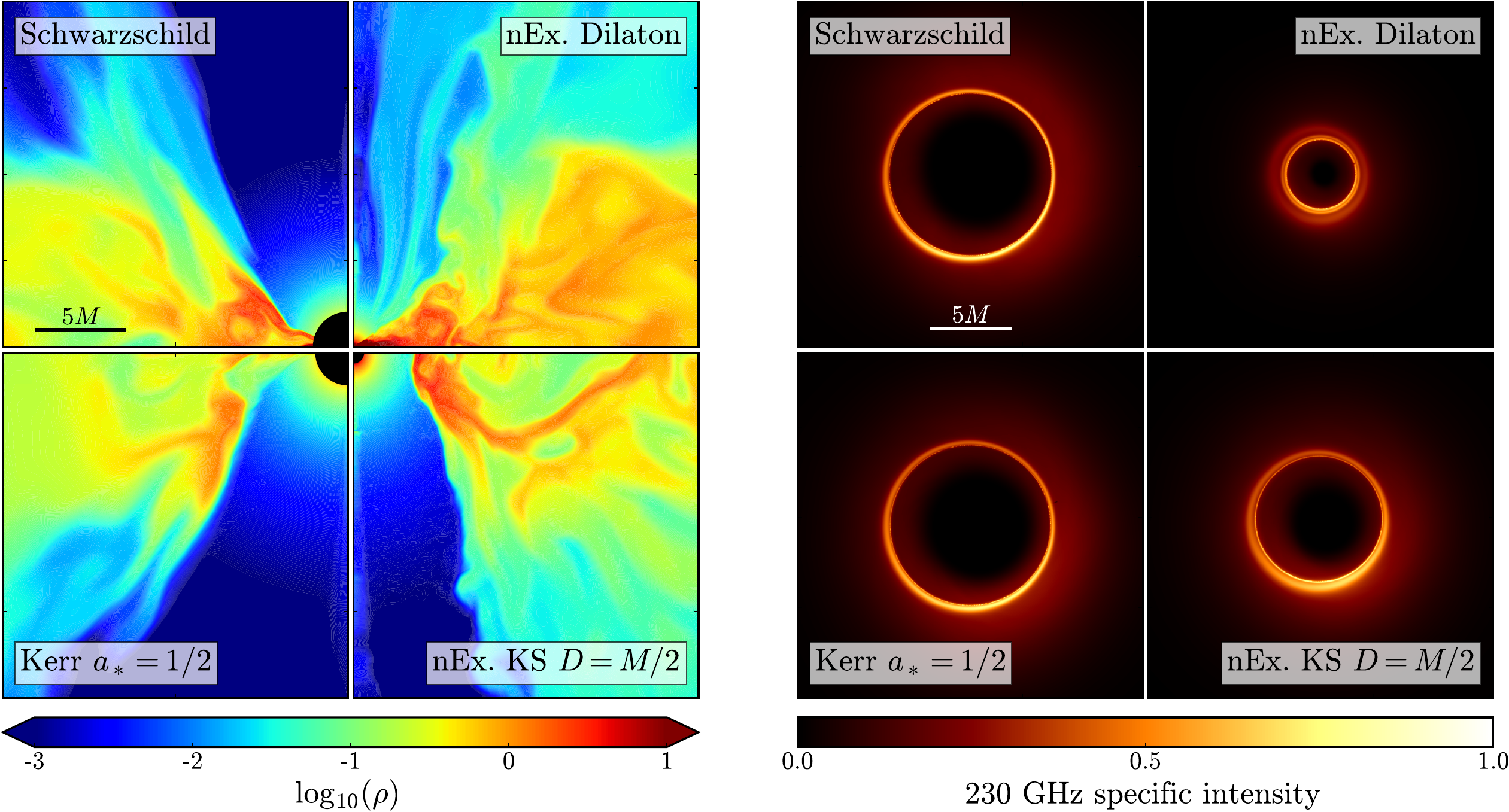}
    \caption{Left: fluid density distribution for four BH models: Schwarzschild, Kerr of BH spin $a_*=a/M=0.5$, near-extremal dilaton of scalar charge $D=0.995M$, and near-extremal dilaton-axion of charge $D=0.5M$ (and $a_*=0.49$). The reduction in the event horizon sizes can be clearly seen for the near-extremal BHs. Right: the corresponding time-averaged 230 GHz (the EHT observing frequency) images for these models are also shown, matched to the M87$^*$ BH parameters \cite{EHTC+2019e}. 
    The bright ring in each image is closely tied to the shadow boundary curve, which is strongly determined by the spacetime geometry and the observer's inclination angle ($163^\circ$). It is again clear to see that there are observable differences in the sizes of these bright rings. The fractional deviation in the mean shadow diameter from the Schwarzschild value, $\delta$, decreases with increasing $D$ or $a_*$ and reaches a minimum of $\delta = 4M/(6\sqrt{3}M) - 1 = -0.615$ for the near-extremal dilaton BH (right top).
    To facilitate comparison, the 2017 EHT observations of M87$^*$ yield a measurement of $\delta=-0.01 \pm 0.17$ \cite{EHTC+2019f, Psaltis+2020, Kocherlakota+2021}, 
    which rules out this BH as a model consistent with M87$^*$ observations \cite{Kocherlakota+2021, Volkel+2021}.
    }
    \label{fig:GRMHD}
\end{figure*}

Furthermore, the observed EHT images have been used to approximately constrain alternative ``non-Kerr'' spacetimes \cite{Psaltis+2020, Kocherlakota+2021, Volkel+2021, EHTC+2022f, Vagnozzi+2022}. However, synthetic images of non-Kerr BHs have largely been constructed using an array of steady-state, semi-analytic accretion models \cite[see, e.g.,][]{Broderick+2014, Shaikh+2019a, Shaikh+2019b, Narayan+2019, Paul+2020, Bauer+2022, Ozel+2022, Younsi+2023, Kocherlakota+2022, Ayzenberg2022, EHTC+2022f}. While such models are extremely insightful, numerical simulations are crucial to capture the dynamic nature of the realistic, turbulent accretion flow. In particular, we expect such time-dependent simulations to provide a sufficiently rich understanding of the temporal and spatial distribution of resolved horizon-scale emission. This would, in turn, provide a more accurate handle on the error in observables such as the shadow sizes of astrophysical objects (see, e.g., \cite{Bardeen1973, Falcke+2000}), which are used to ``measure'' their spacetime geometry \cite{EHTC+2019f, Psaltis+2020, Kocherlakota+2021, EHTC+2022f}. 

Indeed, recent non-Kerr GRMHD simulations have shown that the dilaton BHs of string theory \cite{Gibbons+1988, Garfinkle+1991} may not be easily distinguishable from the Kerr BHs of GR \cite{Mizuno+2018, Fromm+2021, Roder+2023}.
Boson star accretion is another prominent example of such GRMHD simulations, where synthetic images may show a central brightness depression, similar to the EHT-observed images, even though these exotic stars lack event horizons as well as photon shells \cite{Olivares+2020, EHTC+2022f}. Other applications include cold gas accretion in parametrized non-Kerr spacetimes \cite{Nampalliwar+2022}, described by the Johannsen-Psaltis (JP) metric \cite{Johannsen+2010, Johannsen2013}. Hot accretion, relevant for EHT sources, in JP BH spacetimes will be reported in a companion paper \cite{Chatterjee+2023b}.

Thus far, hot accretion in non-Kerr spacetimes has only been explored for the case of nonspinning BHs and nonspinning boson stars. Since astrophysical compact objects are expected to possess nonzero angular momentum in general, this is an important factor to consider. We perform GRMHD simulations in spinning non-Kerr BHs and touch on the corresponding synthetic images below. Moreover, nonzero spin is crucial for the launching of powerful relativistic ``jets'' of magnetized plasma, which is the physical phenomenon that will form our primary focus. 

Jets serve as efficient conduits for transporting vast amounts of energy across cosmic scales and are intimately linked to processes such as star formation and the regulation of galactic dynamics, providing crucial feedback mechanisms \cite{Harrison+2018}.
Jets also accelerate particles to extremely high energies and are thought to be responsible for TeV emission, cosmic rays and gamma-ray flares, thereby offering unique insights into the microphysics of compact objects and their environments \cite[for a comprehensive review, see Ref.][]{Blandford+2019}.

The Blandford-Znajek mechanism \cite[BZ;][]{Blandford+1977} is the most plausible mechanism for powering astrophysical jets. This is an electromagnetic Penrose process \cite{Penrose1969}, i.e., energy is extracted from the BH, causing it to spin down. Thus, understanding how observable jets are powered in general spinning BH spacetimes can provide insight into fundamental aspects of Penrose processes. The BZ mechanism provides a prediction for the jet power, $P_{\rm BZ}$:
\begin{equation} \label{eq:etaBZ}
    P_{\rm BZ}=\eta_{\rm BZ} \dot{M}c^2= \left (\frac{k}{4\pi}\OmegaH^2\phibh^2\right )\dot{M}c^2 \,,
\end{equation}
where $\eta_{\rm BZ}$, $\dot{M}$, $c$, $\phibh$ and $\OmegaH$ are the predicted total outflow efficiency, the mass accretion rate, the speed of light, the dimensionless horizon magnetic flux, and the horizon angular velocity respectively \eqref{eqn:GRMHDeqns}. The proportionality constant $k$ depends on the geometry of the poloidal magnetic field, varying from 0.044 for parabolic to 0.054 for purely radial field lines \citep{Tchekhovskoy+2010}.
While any direct observational evidence of the BZ process is still lacking \cite{Chael:2023}, GRMHD simulations of Kerr BHs have been utilised to verify the BZ jet launching mechanism \citep[e.g.,][]{Komissarov:2001, McKinney+2004,Tchekhovskoy+2010,Narayan+2022} and to constrain the jet power in M87 \citep{EHTC+2019e}.
However, energy extraction mechanisms in spinning, non-Kerr spacetimes have not been explored with GRMHD simulations.

Another goal of this work is to understand, more concretely, the relationship between the BH angular momentum and the launching of outflows, using GRMHD simulations of accretion onto spinning, non-Kerr BHs.
As per Eqn.~\ref{eq:etaBZ}, accretion flows with the maximum possible horizon magnetic flux, or magnetically-arrested disks (hereafter, ``MADs''), exhibit the most powerful jets \citep[e.g.,][]{Tchekhovskoy+2011} for a given $\dot{M}$ and BH horizon angular frequency.

Here we will focus on time-averaged quantities, providing a representative description of our accretion models. We note that while MADs also display transient features frequently such as expulsions of vertical magnetic fields, and the study of these features in non-Kerr spacetimes is an interesting topic in itself, this is beyond the scope of this work. Such eruption events could have possible links to the origin of multiwavelength flares, going up to TeV energies \cite{Dexter+2020,Chatterjee+2021,Porth+2021,Ripperda+2022}, often associated with supermassive BHs.

This work is organized as follows. In Section~\ref{sec:methods} we touch on the Einstein-Maxwell-dilaton-axion theory that provides the background spacetime metric that is used in our GRMHD simulations. Further details can be found in Appendix \ref{sec:EMda_Theory}. In the following section, we discuss the implications of our simulations for horizon-scale images. Sections~\ref{sec:Dilaton_Axion_BH} and \ref{sec:Dilaton_BH} present our main results on energy extraction in non-Kerr spacetimes. In section~\ref{sec:Conclusions} we provide concluding remarks.

\section{Background \& Methodology}
\label{sec:methods}
We first introduce the Einstein-Maxwell-dilaton-axion (EMda) theory that is presently of interest and touch on the differences between its BH solutions and their general-relativistic counterparts.
This EMda theory arises naturally as the low-energy effective limit of the heterotic string (see, e.g., Ch.~8 of Ref.~\cite{Polchinski2007}), and possesses several novel features.
First, it postulates the existence of additional fundamental fields in nature, namely the dilaton scalar and the axion field.
Second, electromagnetism is no longer ``minimally-coupled'' to gravity, as in GR.
While the former is still described by the Maxwell Lagrangian and the latter by the Einstein-Hilbert Lagrangian, their interaction is mediated by the dilaton field.
This can lead to a violation of the weak equivalence principle (see, e.g., \cite{Magueijo2003}).
Third, while the time-invariant (stationary) vacuum BH solutions of both this EMda theory and of GR are given by the Kerr metric, other canonical nonvacuum BH solutions differ due to the differences between the two actions \cite{Plebanski+1976, Garcia+1995}. 

We also present a comparison between the simplest nonvacuum, nonspinning BHs of these two theories, namely the Gibbons-Maeda-Garfinkle-Horowitz-Strominger (GMGHS; \cite{Gibbons+1988, Garfinkle+1991}) BH in this EMda theory and the more familiar Reissner-Nordstr{\"o}m (RN) BH in GR (see, e.g., \cite{Poisson2004}), as well as their spinning versions, the Kerr-Sen \cite[KS;][]{Sen1992} and the Kerr-Newman \cite[KN;][]{Newman+1965b} metrics, in the Appendix. While all of these BHs are electromagnetically charged, to highlight the fields that appear in the EMda counterparts but not in the GR ones, and for simplicity, we will henceforth refer to the GMGHS and the KS BHs as the dilaton and the dilaton-axion BHs respectively. 

\begin{figure}
    \centering
    \includegraphics[width=\columnwidth]{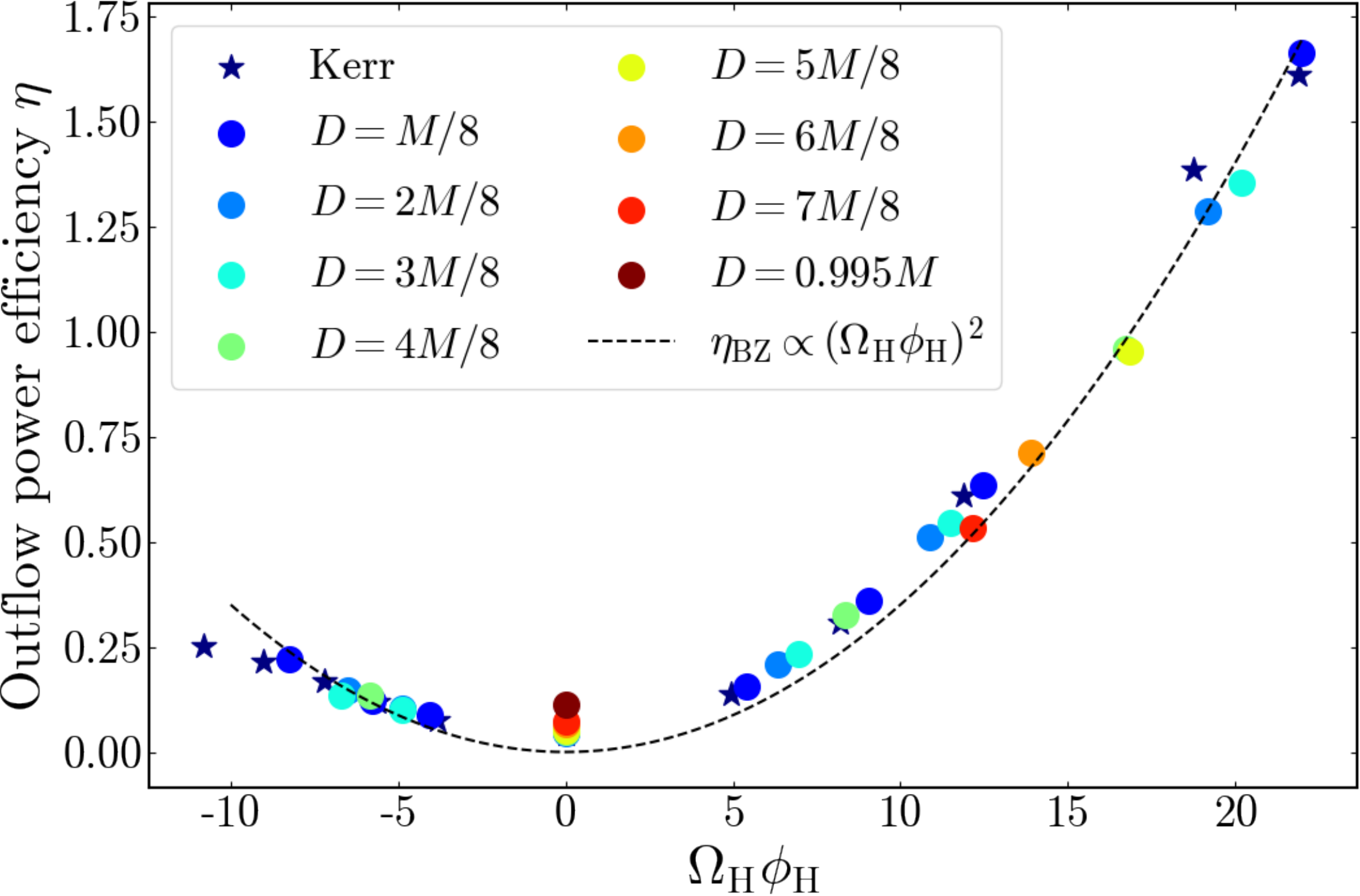}
    \caption{The Blandford-Znajek \citep[BZ][]{Blandford+1977} mechanism predicts the correct jet power even for non general-relativistic BHs.
    We find that the total outflow power $\eta$ matches the predicted BZ power $\eta_{\rm BZ}=(k/4\pi)(\Omega_{\rm H}\phi_{\rm H})^2$ (see eq. \ref{eq:etaBZ}) well when the jets dominate the output power (i.e., for large values of $\Omega$).
    At lower values of $\OmegaH$, $\etaout \gtrsim \eta_{\rm BZ}$ since disk winds contribute to the outflowing power as well (especially for the dilaton models). 
    }
    \label{fig:BZ}
\end{figure}

It is worth emphasizing that astrophysical BHs are not expected to be electromagnetically charged \cite{Blandford+1977,Bransgrove:2021}, even though there may be temporary charge on formation \cite[e.g.,][]{Zilhao+2014, Nathanail+2017}). The dilaton-axion spacetimes we study here should be viewed as popular (due to their connection to string theory) examples of non-GR BHs that are used to explore potential observational signatures of deviations from GR. For simplicity, in this work, we only model the influence of the modified spacetime metric on accretion dynamics \cite[similar to other non-GR BH simulations;][]{Mizuno+2018, Fromm+2021, Roder+2023}, leaving the impact of the background electromagnetic field for future work. We also assume, reasonably, that the dilaton and the axion do not couple to the accretion fluid.

\textit{General-relativistic magnetohydrodynamics simulations}.
We use the GPU-accelerated code \texttt{H-AMR} \citep{Liska+2022} which solves the GRMHD equations in a fixed background spacetime, to perform these high-resolution, three-dimensional simulations. $\texttt{H-AMR}$ requires as input the BH metric in horizon-penetrating coordinates. The desired ``spherical-ingoing Kerr-Schild'' form (with coordinates $t$, $r$, $\vartheta$, $\varphi$) of the general spherically-symmetric and of the axisymmetric metrics can be found in Ref. \cite{Kocherlakota+2023a}, and are given explicitly by eqns.~\eqref{eq:Static_Metric} and \eqref{eq:Stationary_AA_Metric_BL_Coords} in the Appendix here. We adopt geometrized units throughout, i.e. $G=c=1$, which reduces the gravitational radius $\rg=GM/c^2$ to the Arnowitt-Deser-Misner \cite[ADM; ][]{Arnowitt+2008} mass $M$ of the spacetime. 

To investigate energy extraction mechanisms, we construct the quantities relevant for eq. \ref{eq:etaBZ} from our simulations as follows. The total outflow efficiency $\eta$, the dimensionless horizon magnetic flux $\phibh$, and the horizon angular velocity $\OmegaH$ are given respectively as
\begin{align}
    \eta :=&\ \frac{P}{\dot{M}}=\frac{\dot{M}-\dot{E}}{\dot{M}} \,, \\
    \Omega_{\rm H} :=&  \ - \left ( \frac{g_{t\varphi}}{g_{\varphi\varphi}} \right )_{r_{\rm H}} = \ \frac{a_*}{2r_{\rm H}} \,,\\
    \phibh :=&\ \frac{\sqrt{4\pi}}{2\sqrt{\dot{M}}}\iint_{r_{\rm H}} |B^{r}| \, \sqrt{-\mathscr{g}}\, d\vartheta \, d\varphi \,, 
    \label{eqn:GRMHDeqns}
\end{align}
where we have the metric determinant $\mathscr{g}=\det{[g_{\mu\nu}]}$, radial magnetic field $B^r$, the shell-integrated mass accretion rate $\dot{M} = -\iint \rho u^r\sqrt{-g} d\vartheta d\varphi$ comprising of the gas density $\rho$ and the radial component of the fluid 4-velocity $u^{\mu}$. The outflow power $P$ is given in terms of the accretion rate and the total energy flux in the radial direction: $\dot{E}=\iint T^r_t \sqrt{-g} d\vartheta d\varphi$, where the $(r, t)$ component of the stress-energy tensor is $T^r_t=(\rho + \gamma_{\rm ad} U_{\rm g}+b^{\mu}b_{\mu})u^ru_t - b^rb_t$. Here, we have the gas adiabatic index $\gamma_{\rm ad}=13/9$ \cite[same as the MAD models in Ref.][]{EHTC+2019e}, internal energy $U_{\rm g}$, and the comoving 4-magnetic field $b^{\mu}$ \cite{Porth+2019}. We calculate $\phibh$ at the Boyer-Lindquist event horizon radius $r_{\rm H}=M-D+\sqrt{(M-D)^2-a^2}$, while the accretion rate and outflow efficiency at $r=5\,\rg$ in order to avoid contamination from density floors (See Appendix \ref{sec:numerics}). We note that for a given value of $D$, there is a maximum possible BH spin magnitude $|a|=M-D$ beyond which the metric describes a spinning naked singularity. In the above, we have also introduced the dimensionless spin $a_* = a/M$. Finally, we note that all reported quantities are in Boyer-Lindquist coordinates and are time-averaged between $20000M-25000M$. Further description of the numerical setup can be found in the Appendix \ref{sec:numerics}.


\begin{figure*}
    \centering
    \includegraphics[width=2.05\columnwidth]{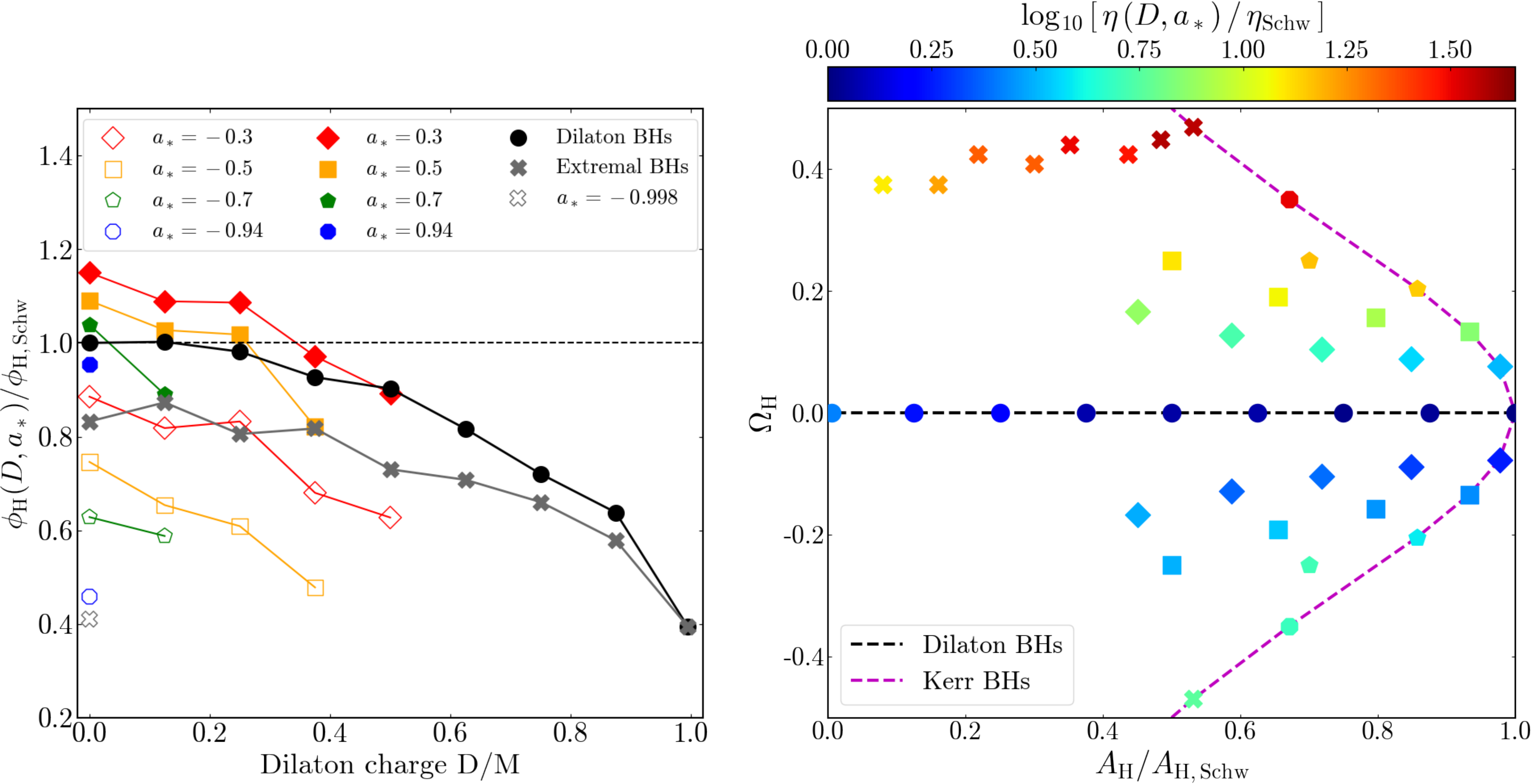}
    \caption{ \textit{Left}: One of the fundamental points of uncertainty for estimating the BZ power output with semi-analytical models of BH accretion is the magnetic flux at the horizon $\phibh$. We find that for magnetically arrested accretion flows, the maximum $\phibh$ depends not only on the BH spin $a_* = a/M$ (as was known for Kerr models, e.g., \citep{Narayan+2022}) but also on the scalar charge $D$. Overall trends show that the saturated magnetic flux decreases with increasing charge for a fixed BH spin, which can be explained by the decreasing BH horizon size. We connect the nonspinning dilaton BHs models using a black line, the extremal BH models with a gray line, and BH models with the same spin by coloured lines. \textit{Right}: Outflow power efficiency $\eta$ changes with both charge and spin. We show $\bar{\eta}=\eta/\eta_{\rm Schw}$ for all Kerr and dilaton-axion GRMHD models as a function of the horizon area $A_{\rm H}$ (normalised to the Schwarzschild value of $16\pi$) and the BH angular frequency $\Omega_{\rm H}$. Higher BH spins result in smaller powerful jets with an increase in charge providing a modest boost in power. Among the models considered in this work, the extremal BH with $D=M/8$ exhibits the largest power.
    }
    \label{fig:KS_phibh}
\end{figure*}


\textit{General-relativistic radiative transfer calculations}.
For the BH imaging calculations was use the general-relativistic radiative transfer (GRRT) code BHOSS \citep{Younsi:2012, Younsi:2016, Younsi:2019}.
The emission mechanism is assumed to be synchrotron radiation from a relativistic Maxwell-J\"uttner distribution of electrons.
The temperature of electrons is calculated from the local ion temperature via $T_{\rm i}/T_{\rm e} := (R_{\rm low} + \beta^2\, R_{\rm high})/(1 + \beta^2)$ \citep{Moscibrodzka:16}, where the plasma-$\beta$ parameter, together with the dimensionless parameters $R_{\rm low}$ and $R_{\rm high}$, modulate the ion-electron coupling.
We choose $R_{\rm low}=1$ and $R_{\rm high}=1$, in order to emphasise the emission from the disk midplane.
All other parameters are chose to reproduce the M87* image properties \citep{EHTC+2019e}, and the time range across which all images are calculated is consistent with the GRMHD simulations, i.e., $20000M-25000M$.

\section{Energy Extraction due to Spin:\\ The Dilaton-Axion Black Hole}
\label{sec:Dilaton_Axion_BH} 

We perform 42 simulations in total: (i) 11 GR Kerr BH simulations of varying spacetime specific angular momentum, $a$, (ii) 8 dilaton BH simulations of varying ADM dilaton/scalar charge, $D$, and (iii) 23 dilaton-axion BHs that are the spinning versions of the dilaton model set. Of these, 8 dilaton-axion models are ``near-extremal'' BHs, i.e., $|a|\approx M-D$. Table \ref{tab:models} in the appendix provides the BH parameters for our GRMHD simulation set.
Figure~\ref{fig:GRMHD} presents a collection of four representative GRMHD snapshots and the corresponding synthetic M87* images which result from their subsequent GRRT post-processing. 

Our first goal is to establish whether the BZ process can explain the outflow power from our models. The strength of the BZ power efficiency lies in its simplicity: $\eta_{\rm BZ}\propto (\OmegaH\phibh)^2$, i.e., it is the square of a single number. This means that the BZ formula can be easily tested for any BH accretion model. Note that this is not the case for extended BZ power models where the outflow efficiency is $\eta_{\rm BZ}\times[1+\mathscr{O}(\OmegaH^2)+\mathscr{O}(\OmegaH^4)]$ \citep{Tchekhovskoy+2010}, thus independently requiring both $\OmegaH$ and $\phibh$ explicitly. Since jets are a direct result of the twisting of magnetic fields due to spin, the jet power we obtain should follow the BZ formula. This is exactly what we see for large values of $\OmegaH\phibh$ in Fig.~\ref{fig:BZ}, where we expect the jet power to dominate over the power from accretion disk outflow (or ``winds''), demonstrating that the BZ mechanism describes the power of jets even when considering these non-GR dilaton-axion BHs \cite[for non-Kerr BHs described by the JP parametrized metric, see Ref.][]{Chatterjee+2023b}.

Note that one of the ingredients of the BZ power formula, namely the horizon magnetic flux $\phibh$, cannot be determined by semi-analytical methods and requires the numerical simulations. From the left panel of Fig.~\ref{fig:KS_phibh}, we see that $\phibh$ depends on not only $a$ \cite[as was found before for Kerr BHs; see Refs][]{Tchekhovskoy+2012, Narayan+2022} but also generally decreases monotonically with increase in $D$. This is because the horizon shrinks in size faster compared to the increase in the magnetic field strength at the horizon. Furthermore, prograde BHs (i.e. where the BH corotates with the accretion flow) exhibit larger $\phibh$ values for the same $|a|$. While the behavior $\phibh$ has not yet been fully explained, it becomes clear from this figure that both changes in $a$ as well as changes in the horizon size can independently play a role. We note also that the gas density distribution and the geometrical height of the accretion flow can also impact $\phibh$ \cite[e.g.,][]{Narayan+2022}. 

To facilitate semi-analytical non-GR BH jet models, we provide a fitting model for the magnetic flux as a function of $a$ and the horizon area ratio $\bar{A}_{\rm H} := A_{\rm H}/A_{\rm H\,, Schw}$,
\begin{align}
\phi_{\rm H\,, fit}(a_*, \bar{A}_{\rm H}) =&\ \phi_{\rm H\,, Schw}\cdot\phi_{\rm a}(a_*)\cdot\phi_{\rm A}(\bar{A}_{\rm H})\,,
\end{align}
where $\phi_{\rm H\,, Schw}=56.05$ is the value for a Schwarzschild BH, $\phi_{\rm a}(a_*) := 1 + 0.4 a_* - 0.354 a_*^2 - 0.177 a_*^3$ measures the impact of spin and $\phi_{\rm A}(\bar{A}_{\rm H}) := 1 + 0.41 (1-\bar{A}_{\rm H}^{1/4}) -1.407 (1-\bar{A}_{\rm H}^{1/2})+0.42 (1-\bar{A}_{\rm H}^{3/4})$ measures the impact of a varying horizon size. The horizon area $A_{\rm H}$ is given in terms of the metric components $g_{\mu \nu}$ as $A_{\rm H} := \iint \sqrt{g_{\vartheta\vartheta}g_{\varphi\varphi}} d\vartheta d\varphi|_{r=r_{\rm H}}$, which in our case yields $A_{\rm H} = 4\pi(r_{\rm H}^2+2Dr_{\rm H}+a^2)$. While $a_*:=a/M$ and $\bar{A}_{\rm H} = A_{\rm H}/(16\pi M^2)$ are not independent variables, our choice to employ these coordinate-invariant, dimensionless quantities here enables comparisons with generic BH spacetimes, including those that do not possess a dilaton or an axion field. 

In addition to $\phibh$, we expect the jet power to vary also as $\propto\OmegaH^2\propto r_{\rm H}^{-2}$ (eq. \ref{eq:etaBZ}, Fig. \ref{fig:BZ}). Thus for the same BH spin, increasing $D$ would reduce $\phibh$ but increase $\OmegaH$. The right-hand plot of Fig.~\ref{fig:KS_phibh} shows this dichotomy in the behavior of $\eta$ very well. Generally, we find that (i) models with the same $a$ exhibit larger $\eta$ with an increase in $D$, (ii) models with the similar $\OmegaH$ show decreasing $\eta$ when reducing the horizon radius (as expected due to the decrease in $\phibh$), and (iii) the highest spinning models show the largest values of $\eta$. 

Furthermore, curiously, since $\phibh$ decreases for $a>0.3M$ (see, e.g., \cite{Narayan+2022}, Fig \ref{fig:KS_phibh}), near-extremal BHs can show a slight increase in $\eta$ for small $D$. In particular, the power for the near-extremal, spinning BH with $D=M/8$ is larger than that for the near-extremal Kerr BH ($a=0.998M$) with $D=0$. Indeed, using our fitting function $\phi_{\rm H, fit}$, we predict that the outflow efficiency, given as $\eta_{\rm BZ, fit}=(k/4\pi)\phi_{\rm H, fit}^2\OmegaH^2$, takes its maximal value of $1.87$ (or $187\%$) for the near-extremal BH with $a=0.8M$ and $D=0.2M$. As $D\approx M-a$, we expect that our results above indicate approximate features of realistic magnetized, hot accretion in actual (non-)spinning naked singularity spacetimes, for the first time. 

Our results show that the introduction of spin to the spacetime plays the primary role in boosting the outflowing power due to the presence of jets. In addition to the Poynting-flux dominated jet, there are also gas outflows that originate from the disk. These ``disk winds'' are typically slow-moving and carry gas away from the disk. BZ-driven jets inject their electromagnetic energy into the surrounding winds. As such, the disk wind power usually increases with increasing BZ power. Despite reaching far smaller powers, winds make a significant contribution to the total outflow power $\eta$ for spins $a\lesssim 0.7M$. Interestingly, the power output $\eta$ for dilaton BHs ($\OmegaH=0$) show the opposite behaviour as jetted BHs with a constant $\OmegaH$, increasing with decreasing horizon area. Evidently, the outflow mechanism is different for dilaton BHs, which we study next.


\begin{figure}
    \centering
    \includegraphics[width=\columnwidth,trim= 60pt 20pt 80pt 80pt, clip]{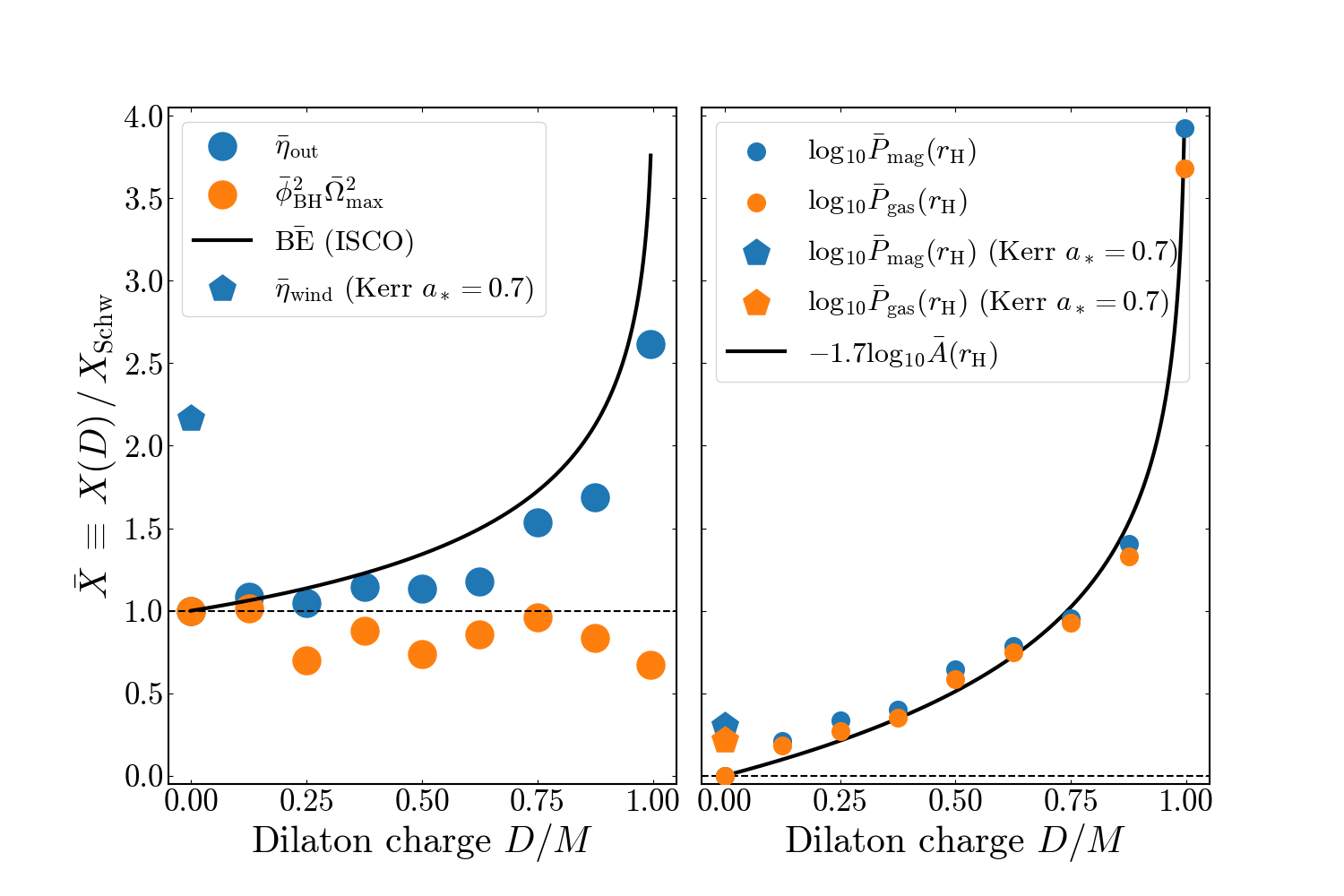}
    \caption{Energy Extraction from nonspinning dilaton BHs: increasing the dilaton charge powers disk winds. We show the dilaton BH quantities relative to their Schwarzschild counterparts $\bar{X}=X(D)/X_{\rm Schw}$. Higher $D$ values exhibit larger outflow power efficiencies ($\eta$), which can be attributed to the large magnetic and thermal pressures ($P_{\rm mag}$ and $P_{\rm gas}$) felt by the gas. We also show the binding energy (BE) since this is a good indicator of the energy released by accretion perhaps to power outflows. For comparison, we also show the corresponding wind outflow efficiency for a Kerr BH of spin $a=0.7M$. 
    }
    \label{fig:GMGHS}
\end{figure}

\section{Energy Extraction due to Curvature:\\ The Zero-Spin Dilaton Black Hole}
\label{sec:Dilaton_BH}

While both dilaton-axion and Kerr BHs become smaller with increasing spin, the nonspinning dilaton metric grants us a new handle (via $D$) to study the impact of a varying BH size on accretion dynamics, while leaving the spin degree-of-freedom unchanged ($a=0$). Equivalently, a systematic exploration of accretion in this spacetime allows us to address how unique signatures of plasma accessing higher curvatures (stronger ``gravitoelectric'' effects; cf. \cite{Ciufolini+1995, Mashhoon+2003}) in the absence of frame-dragging (no ``gravitomagnetic'' effects) can appear in BH images. In our most extreme case ($D=0.995M$), the dilaton BH horizon is located at $r_{\rm H} = 0.01M$ and has a horizon area that is 200 times smaller than a Schwarzschild BH with same $M$. 

As we remarked in the previous section, the outflow power for dilaton BHs grows with increasing $D$ and can exceed the total outflow power ($\eta=\eta_{\rm jet}+\eta_{\rm wind}$) exhibited by slowly rotating retrograde Kerr BHs with jets such as $a=-0.3M$ (Fig.~\ref{fig:KS_phibh}). Indeed, it can even be larger than the ``wind-only'' outflow power seen in Kerr $a=0.7M$ BHs (Fig.~\ref{fig:GMGHS}). While the wind launching mechanism cannot be the BZ process (since $\OmegaH=0$), we first construct a pseudo-BZ-like power prescription, $\propto\phibh^2\Omega_{\rm max}^2$ that may be taken as an upper limit of the well-known Blandford-Payne \cite{Blandford+1982} wind power $\eta_{\rm BP}\propto \phi^2\Omega/r$ \cite{Livio+1999}, where $\Omega$ is the gas angular velocity. The maximum $\Omega$ in jetted BHs is at the horizon, while for nonspinning BHs, $\Omega_{\rm max}$ occurs near the innermost circular stable orbit radius \cite[ISCO;][]{Shakura+1973,Novikov+1973}. 

The substantial mismatch between $\eta$ and our pseudo-BZ power seen in Fig.~\ref{fig:GMGHS}a suggests that the outflow launching mechanism may not be related to the gas rotational energy. Other than rotation, outward forces could be generated via large outward pressure gradients or magnetic tension due to jets or magnetic explosions often seen in MAD accretion flows \cite{Chatterjee+2022, Begelman+2022}. Indeed, we see that the magnetic and thermal pressure rapidly increase for dilaton BHs. In advection-dominated accretion flows \cite[][]{Yuan+2014}, as is the case for our models, the energy released due to viscous dissipation, i.e., the binding energy (BE), does not radiate out and instead increases the thermal pressure, generating a hot accretion flow, while the rest may be released in the form of outflows. In our case, some of the BE can also transfer to the disk magnetic energy. Therefore, as $D$ increases, both the magnetic and thermal pressures increase as $A_{\rm H}^{-1.7}$ (Fig.~\ref{fig:GMGHS}b). We show the binding energy of the innermost stable circular orbit (ISCO) as an indicator of the available energy in the disk given a particular BH metric solution and find that it strongly correlates with the outflow power (Fig.~\ref{fig:GMGHS}a). This verifies that the release of energy due to accretion onto deeper curvatures power disk winds from dilaton BHs. 

\section{Summary}
\label{sec:Conclusions}

We perform an extensive suite of 42 high-resolution, fully 3D GRMHD simulations of hot accretion flows onto non-Kerr BHs using the Kerr-Sen or, equivalently, the dilaton-axion metric. These not only include the first simulations in spinning non-Kerr BH spacetimes but also the first simulations in near-extremal non-Kerr BH spacetimes. These are particularly interesting because they enable connecting potentially observable large-scale MHD dynamics to fundamental solutions for stringy BHs. Further, the horizon radius can shrink significantly as the dilaton charge increases, resulting in strikingly smaller shadow sizes, which are already measurable from the EHT images. This enables us to test jet and accretion physics in extreme curvatures beyond what is possible with the standard Kerr metric. 

Our goal in this work is to understand how the energy can be extracted from the BH to power relativistic jets. We find that the Blandford-Znajek (BZ) mechanism is active even in non-general-relativistic BH spacetimes. Further, the non-spinning dilaton BH simulations allow to us study the nature of outflows in the absence of the BZ process, which depends on frame-dragging, more extensively in addition to the Schwarzschild simulation. In these BHs, we find that energy extraction works via disk winds powered by the available binding energy of the accretion flow. Smaller horizon sizes show stronger curvatures that result in the gas achieving higher binding energies and therefore, more powerful winds.

While this paper focuses on outflow properties, much of the rich accretion physics in these dilaton-axion simulations remains to be explored. Of particular interest could be the dominant radial nature of the gas inflow as well as the evolution of magnetic explosions that are triggered when the BH magnetosphere becomes over-saturated in these MAD flows. Both of these effects would significantly alter the variability of the horizon-scale images as well as the time evolution of high-energy flaring events. With advances in very-long-baseline-interferometry (VLBI) such as the EHT and the possibility of space-based VLBI, measurements of BH spin and deviations in shadow shape from Schwarzschild or Kerr BHs, such as those seen in Fig.~\ref{fig:GRMHD}, could become a reality for constraining string theory BHs.


\begin{acknowledgments}
KC, PK, and RN acknowledge support from grants from the Gordon and Betty Moore Foundation and the John Templeton Foundation to the Black Hole Initiative at Harvard University, and from NSF award OISE-1743747. ZY acknowledges support from a UK Research and Innovation (UKRI) Stephen Hawking Fellowship. This research was enabled by support provided by a INCITE program award PHY129, using resources from the Oak Ridge Leadership Computing Facility, Summit, which is a US Department of Energy office of Science User Facility supported under contract DE-AC05- 00OR22725, as well as Calcul Quebec (http://www.calculquebec.ca) and Compute Canada (http://www.computecanada.ca).
This work has made use of NASA's Astrophysics Data System (ADS).
\end{acknowledgments}


\bibliography{v3-Refs-KS_GRMHD}


\begin{appendix}

\section{The Einstein-Maxwell-dilaton-axion Theory and Associated Black Hole Solutions}
\label{sec:EMda_Theory}

Here we first briefly introduce the Einstein-Maxwell-dilaton-axion (EMda) theory that is presently of interest and touch on the differences between its black (BH) solutions from their general relativistic counterparts. This EMda theory arises naturally as the low-energy effective limit of the heterotic string (see, e.g., Ch. 8 of Ref. \cite{Polchinski2007}), and possesses several novel features. First, it predicts the existence of additional fundamental fields in nature, namely the dilaton scalar and the axion field. Second, electromagnetism is no longer ``minimally-coupled'' to gravity, as in GR. While the former is still described by the Maxwell Lagrangian and the latter by the Einstein-Hilbert Lagrangian, their interaction is mediated by the dilaton. This can lead to a violation of the weak equivalence principle (see, e.g., \cite{Magueijo2003}). Third, while the time-invariant (stationary) vacuum BH solutions of this EMda theory as well as of GR are given by the Kerr metric, other canonical nonvacuum BH solutions differ due to the differences between the two actions \cite{Plebanski+1976, Garcia+1995}. 

The simplest examples of analogous spacetimes in these two theories which exhibit significant differences correspond to their electromagnetically charged BHs. 
In particular, the nonspinning, spherically-symmetric, charged BHs are given by the Gibbons-Maeda-Garfinkle-Horowitz-Strominger (GMGHS; \cite{Gibbons+1988, Garfinkle+1991}) metric in this EMda theory and by the more familiar Reissner-Nordstr{\"o}m (RN) metric in GR (see, e.g., \cite{Poisson2004}). Both metrics can be written in the form,
\begin{equation} \label{eq:Static_Metric}
\mathrm{d}s^2 = -f\mathrm{d}t^2 + \frac{\mathrm{d}r^2}{f} + R^2\mathrm{d}\Omega_2^2\,,
\end{equation}
where the two metric functions $f$ and $R$ depend only on $r$. For the GMGHS spacetime, these take values,
\begin{equation} \label{eq:GMGHS_Metric_Functions}
f(r) = \left(1 - \frac{2M}{r + 2|D|}\right)\,,\ 
R^2(r) = r\left(r + 2|D|\right)\,,
\end{equation}
whereas for the RN spacetime, these are given as,
\begin{equation} \label{eq:RN_Metric_Functions}
f_{\mathrm{RN}}(r) = \left(1 - \frac{2M}{r} + \frac{Q^2}{r^2}\right)\,,\ 
R_{\mathrm{RN}}^2(r) = r^2\,.
\end{equation}
In the above, $M$ denotes the Arnowitt-Deser-Misner (ADM; \cite{Arnowitt+2008}) mass of the spacetime, $|D|$ the (absolute) ADM dilaton/scalar charge, and $Q$ its ADM electromagnetic charge. For a magnetically-charged GMGHS BH, $D$ is negative and is positive for the electrically-charged one. In both of these cases, an axion field is absent and its electromagnetic charge $Q$ cannot be freely specified; Instead, it is given as $Q=\pm\sqrt{2M|D|}$ (cf. eq. 10 of Ref. \cite{Garfinkle+1991}). Note that the ``original'' form of the metric given in eq. 6 there is written in a different radial coordinate. We can put the metric above \eqref{eq:GMGHS_Metric_Functions} into its original form by making the replacement $r \mapsto r-2|D| = r - Q^2/M$. We prefer to use the coordinate system above \eqref{eq:Static_Metric} since the curvature singularity is then located at $r=0$. 

The spinning, axisymmetric, charged BHs of this EMda theory and of GR are described by the Kerr-Sen (KS; \cite{Sen1992}) and the Kerr-Newman (KN; \cite{Newman+1965b}) metrics respectively. Both metrics, in Boyer-Lindquist (BL) coordinates $x^\mu=(t,r,\vartheta,\varphi)$, take the following general form (cf. \cite{Kocherlakota+2023a, Xavier+2020}),
\begin{align} \label{eq:Stationary_AA_Metric_BL_Coords}
\mathrm{d}s^2 =&\ g_{\mu\nu}\mathrm{d}x^\mu\mathrm{d}x^\nu =  
-\left(1-\frac{2 F}{\Sigma}\right)\mathrm{d}t^2 
-2\frac{2 F}{\Sigma}a\sin^2{\vartheta}~\mathrm{d}t\mathrm{d}\mathrm{\varphi}
\nonumber \\
&\ \quad\quad\quad
+ \frac{\Sigma}{\Delta}\mathrm{d}r^2 + \Sigma~\mathrm{d}\vartheta^2 +\frac{\Pi}{\Sigma}\sin^2{\vartheta}~\mathrm{d}\varphi^2\,,
\end{align}
where the spinning BH metric functions $F, \Delta, \Sigma$, and $\Pi$ are given in terms of the nonspinning BH metric functions $f$ and $R$ above \eqref{eq:Static_Metric} simply as \cite{Kocherlakota+2023a},
\begin{align} \label{eq:AA_Metric_Functions}
F(r) =&\ (1 - f)R^2/2\,, \\
\Delta(r) =&\ f R^2 + a^2\,, \nonumber \\
\Sigma(r, \vartheta) =&\ R^2 + a^2\cos^2{\vartheta}\,, \nonumber \\
\Pi(r, \vartheta) =&\ \left(R^2 + a^2\right)^2 - \Delta a^2\sin^2{\vartheta}\,. \nonumber 
\end{align}
Here $a=J/M$ is the specific angular momentum as usual, where $J$ the ADM angular momentum of the spacetime. We note here for convenience that the determinant of the metric $\mathscr{g}:=\det{[g_{\mu\nu}]}$ and of the $t\varphi-$sector $\mathscr{g}_{t\varphi} := g_{tt}g_{\varphi\varphi} - g_{t\varphi}^2$ are given as,
\begin{equation}
\mathscr{g} = -\Sigma^2\sin^2{\vartheta}\quad \mathrm{and}\quad \mathscr{g}_{t\varphi} = -\Delta\sin^2{\vartheta}\,.
\end{equation}

To obtain the KS and the KN metrics specifically, we use the metric functions given in eq. \ref{eq:GMGHS_Metric_Functions} and in eq. \ref{eq:RN_Metric_Functions} respectively in eqs.  \ref{eq:AA_Metric_Functions} and \ref{eq:Stationary_AA_Metric_BL_Coords} above. Despite the KS spacetime containing dilaton and axion fields in addition to the electromagnetic one, similar to the case of the GMGHS BH above, it remains a three-parameter (${M, D, a}$) solution \cite{Sen1992}. In the limit of vanishing spin ($a\rightarrow 0$), the KS and KN (spinning/stationary) metrics reduce to the GMGHS and RN (nonspinning/static) ones respectively. Furthermore, as noted above, in the limit of vanishing charge ($D$ or $Q$), it is easy to see that we obtain the Kerr metric in either case.

The horizons are always present at BL locations, $r=r_{\pm}$, where $g^{rr} = 0$, i.e., at $\Delta(r)=0$. In the main text, the outer horizon is indicated by $r_{\rm H}\equiv r_+$. For the KS BH, these are at 
\begin{equation} \label{eq:KS_Horizons}
r_{\pm} = M - |D| \pm \sqrt{\left(M - |D|\right)^2 - a^2}\,,
\end{equation}
whereas for the KN BH, one obtains,
\begin{equation} \label{eq:KN_Horizons}
r_{\pm\,, \mathrm{KN}} = M \pm \sqrt{M^2 - Q^2 - a^2}\,.
\end{equation}
For a fixed BH mass $M$, extremal BHs, for which $r_+ = r_-$, in either case (KS or KN), form a single-parameter family of metrics. For the extremal KS BHs, $r_+ = a$, and we can write,
\begin{equation} \label{eq:KS_Extremal_Parameters}
|D_{\mathrm{ex}}(a)| = M - a\quad \mathrm{or}\quad \ a_{\mathrm{ex}}(D) = M - |D|\,,
\end{equation}
whereas for the extremal KN BHs, $r_+ = M$, and we have,
\begin{equation} \label{eq:KN_Extremal_Parameters}
Q_{\mathrm{ex\,, KN}}(a) = \pm\sqrt{M^2 - a^2}\quad \mathrm{or}\quad \ a_{\mathrm{ex\,, KN}}(Q) = \sqrt{M^2 - Q^2}\,.
\end{equation}
Thus, the KS and the KN metrics describe BH spacetimes when $a+|D| \leq M$ and for $a^2 + Q^2 \leq M^2$ respectively. While the differences between the ``stringy'' KS BHs and the general-relativistic KN BHs are clear to see from the discussion above, we highlight one particularly striking difference (see also \cite{Garfinkle+1991}). The nonspinning GMGHS BH does not actually possess an inner (Cauchy) horizon, i.e., $r_-(a=0) = 0$ \eqref{eq:KS_Horizons}, unlike its general-relativistic counterpart. Therefore, a GMGHS BH has a spacelike singularity at its center, similar to a Schwarzschild BH, whereas an RN BH has a timelike one instead (see, e.g., \cite{Poisson2004}). Furthermore, not only does the GMGHS metric not admit an extremal GMGHS BH at $|D_{\mathrm{ex}}(a=0)| = M$ but, instead, it describes a naked singularity spacetime since $r_+ = 0$ then. Thus, even in the absence of an axion field, simply due to the presence of the dilaton, there are significant differences between the BHs from these two theories.


\begin{table}
\centering
\vspace{10pt}
\renewcommand{\arraystretch}{1.3}
\begin{tabular}{| c | c | c |}
\hline
Metric & Model  & Black Hole Spin $a_*$\\
& ($D/M$) & \\
\hline
Kerr & 0.0 & -0.998, -0.94, -0.7, -0.5, -0.3, 0.0, \\ && 0.3, 0.5, 0.7, 0.94, 0.998 \\
Dilaton-Axion & 0.125 & -0.7, -0.5, -0.3, 0.0, 0.3, 0.5, 0.7, 0.87 \\
Dilaton-Axion & 0.250 &  -0.5, -0.3, 0.0, 0.3, 0.5, 0.74 \\
Dilaton-Axion & 0.375 &  -0.5, -0.3, 0.0, 0.3, 0.5, 0.62\\
Dilaton-Axion & 0.500 &  -0.3, 0.0, 0.3, 0.49\\
Dilaton-Axion & 0.625 &  0.0, 0.37\\
Dilaton-Axion & 0.750 &  0.0, 0.24\\
Dilaton-Axion & 0.875 &  0.0, 0.12\\
Dilaton-Axion & 0.995 &  0.0\\

\hline
\end{tabular}
\caption{List of GRMHD simulations in this work.} 
\label{tab:models}
\end{table}

\section{Numerical Implementation}
\label{sec:numerics}

Our simulations have an effective grid resolution of $N_r\times N_{\vartheta}\times N_{\varphi}=384\times 240\times256$. The grid extends from $r\in (0.85~r_{\rm H},10^4~M)$, $\vartheta\in (0,\pi)$ and $\varphi\in (0,2\pi)$. For the near-extremal dilaton-axion BHs, our horizon radius $r_{\rm H}$ reduces significantly, requiring additional cells in the radial direction. For e.g., the $D=0.995~M$ dilaton model with $r_{\rm H}=0.005~M$ has an effective radial grid resolution of $N_r=1044$. To aid computational speed for these cases, we derefine the grid by a factor of $2\times2\times2$ within $4~M$. We adopt outflowing radial boundary conditions (BCs), transmissive polar BCs and periodic BCs in the $\varphi-$direction \citep{Liska+2018, Liska+2022}. We initialise the disk in the form of an equilibrium hydrodynamic torus \citep{Fishbone+1976} around a central BH. We set up a standard MAD magnetic field configuration \citep[e.g.,][]{Chatterjee+2022}. For our gas thermodynamics, we assume an ideal gas equation of state with an adiabatic index of $13/9$. The magnetic field strength is normalised by setting the initial maximum gas-to-magnetic pressure ratio to 100. For tackling the evacuated region in the jet funnel, we adopt the density floor injection scheme of \citet{Ressler+2017} when the magnetisation exceeds 20. Table~\ref{tab:models} provides the list of simulation models explored in this work.

\end{appendix}

\end{document}